\documentclass[twocolumn]{aastex7}

\usepackage{array} 
\usepackage{amsfonts}
\usepackage{amssymb} 
\usepackage{amsmath}
\usepackage{graphicx}
\usepackage{enumerate}
\usepackage{verbatim}
\usepackage{color}
\usepackage{mathrsfs}
\usepackage{comment}
\usepackage{multirow} 
\usepackage{diagbox}
\bibpunct{(}{)}{;}{a}{}{,} 
\usepackage{url}
\usepackage{hyperref}
\usepackage{bm}
\usepackage{enumerate}
\usepackage{mathrsfs}
\usepackage{comment}

\def\aj{{AJ}}                   
\def\apj{{ApJ}}                 
\def\apjl{{ApJ}}                
\def\apjs{{ApJS}}               
\def\aap{{A\&A}}                
\def\apss{{Ap\&SS}}          
\def\mnras{{MNRAS}}             
\def\nat{{Nature}}              

\def\psj{{PSJ}}

\newcommand{\vl}{2005~VL$_1$}
\newcommand{\of}{\texttt{orbit\_finder}}

\begin{document}

\title{``Dark comet" 2005 VL$_1$ is unlikely to be the lost Soviet-era probe Venera 2}
\author[orcid=0000-0001-6948-4259,sname='Spada']{Federico Spada}
\affiliation{Universit\`a di Catania, Dipartimento di Fisica e Astronomia}
\email[show]{federico.spada@dfa.unict.it}

\begin{abstract}
\citet{Loeb_Cloete:2025} intriguingly suggest that the near-Earth object 2005 VL$_1$ could be the lost Soviet probe Venera 2.
Here I evaluate the plausibility of such a claim against the available data.
I have re-determined the orbit of 2005 VL$_1$ (including a non-gravitational acceleration component) using the astrometric observations retrieved from the Minor Planet Center (MPC) database.
By propagating the orbit of 2005 VL$_1$ over the period of the Venera 2 mission, I compare this object’s distance from the Earth and from Venus at the times of the probe’s launch and flyby with Venus, respectively.
My analysis, which takes into account realistic uncertainties on both the orbit of 2005 VL$_1$ and the position of Venera 2, decisively rules out the proposed identification.
My approach relies entirely on open-source software and publicly available data, and could represent a viable method to assess similar claims in the future.
\end{abstract}

\keywords{astrodynamics --- near-Earth objects --- planetary probes --- orbit determination}

\section{Introduction}

More than sixty years have passed since the dawn of the space era. 
Over the last few decades, the chance of observing man-made artifacts when looking at the night sky went from negligible to high, and nowadays poses a serious threat to ground-based observational astronomy \citep[e.g.,][]{Lawler_ea:2022}. 
In parallel to the rise of artificial satellites, the number and sensitivity of observational facilities dedicated to the search of solar system minor bodies has also increased, in an effort to strengthen our planetary defense capabilities.
For these reasons, the possibility that an observed celestial body is a man-made object should not be discounted a priori, and similar identifications have been convincingly carried out in the past \citep[e.g.,][]{Jorgensen_ea:2003}.

Most recently, \citet{Loeb_Cloete:2025} have suggested that the near-Earth asteroid \vl{} could in fact be the Soviet-era probe Venera 2, which was lost in February 1966 shortly after its Venus flyby.
Having a significant non-gravitational acceleration compatible with outgassing of volatiles, but no detectable coma (somewhat analogously to the interstellar object 1I/'Oumuamua) \vl{} has been classified as a ``dark comet" by \citet{Seligman_ea:2023}.
On the basis of their tentative identification with Venera 2, and given the expected area-to-mass ratio of the probe, \citet{Loeb_Cloete:2025} propose an alternative explanation of the detected non-gravitational acceleration of \vl{} in terms of solar radiation pressure. 

Using the observational constraints on the orbit of \vl{} and the available information on the trajectory of Venera 2, I show that the proposed identification of these two objects is unlikely to be correct (see also \citealt{Deen2025}; \citealt{McDowell2025}).

\section{Data and orbit comparison}

The data available for \vl{} consists of astrometric position measurements at $79$ epochs between 2005 Nov 04 and 2021 Dec 04.

To constrain the trajectory of Venera 2, I rely on the following notable events: i) the probe was launched on 1965 Nov 12 at 04:46 UTC, and about one hour later injected into heliocentric orbit starting from an altitude $\approx 300$ km; ii) Venus flyby occurred on 1966 Feb 27 at 02:52 UTC, at a range of approximately $24000$ km (\citealt{siddiqi2018}; \citealt{McDowell2025}). 

I have independently re-determined the orbit of \vl{} from the available astrometry by performing a standard orbit determination procedure (see \citealt{Spada:2023} and references therein) using the \of{} code.
This code is written in Python and available on GitHub \citep{orbitfinder}; it implements the weighting scheme of \citet{Veres_ea:2017}, and the bias correction scheme of \citet{Eggl_ea:2020}; orbit propagation is performed using the ASSIST extension of the N-body code REBOUND (\citealt{Holman_ea:2023}, and \citealt{Rein_Liu:2012}, respectively).

Besides purely gravitational effects (see \citealt{Holman_ea:2023} for details), my orbital fit of \vl{} includes a non-gravitational acceleration formulated according to the \citet{Marsden_ea:1973} parametric model, with $g(r) = (1\, {\rm au}/r)^2$. 

My fit for the epoch 2025 May 05 utilizes 78 of the 79 available epochs, and has a final RMS error of $0.49$ arc seconds; the classical orbital parameters in the J2000 ecliptic heliocentric frame are: $a = 0.8911689 \pm 1.23 \cdot 10^{-7}$ au, $e =  0.2246163 \pm 1.75 \cdot 10^{-7}$, $I = 0.2356904 \pm 3.07 \cdot 10^{-6}$, $\Omega =  37.8229215 \pm  2.05 \cdot 10^{-3}$, $\omega =   227.8469644 \pm  2.06 \cdot 10^{-3}$, $M =   175.4168105 \pm 1.93 \cdot 10^{-5}$ (all angles expressed in degrees), while the non-gravitational acceleration parameters are $A_1 =    -7.99 \cdot 10^{-10} \pm  4.57 \cdot 10^{-11}$, $A_2 =    -8.17 \cdot 10^{-13} \pm  1.93 \cdot 10^{-14}$, $A_3 =    -2.44 \cdot 10^{-11} \pm  3.45 \cdot 10^{-12}$ (in au day$^{-2}$). 
This orbital fit, with its uncertainty, is my baseline for comparison with the Venera 2 trajectory constraints, shown in Figure 1. 

\begin{figure*}
\begin{center}
\includegraphics[width=0.9\textwidth]{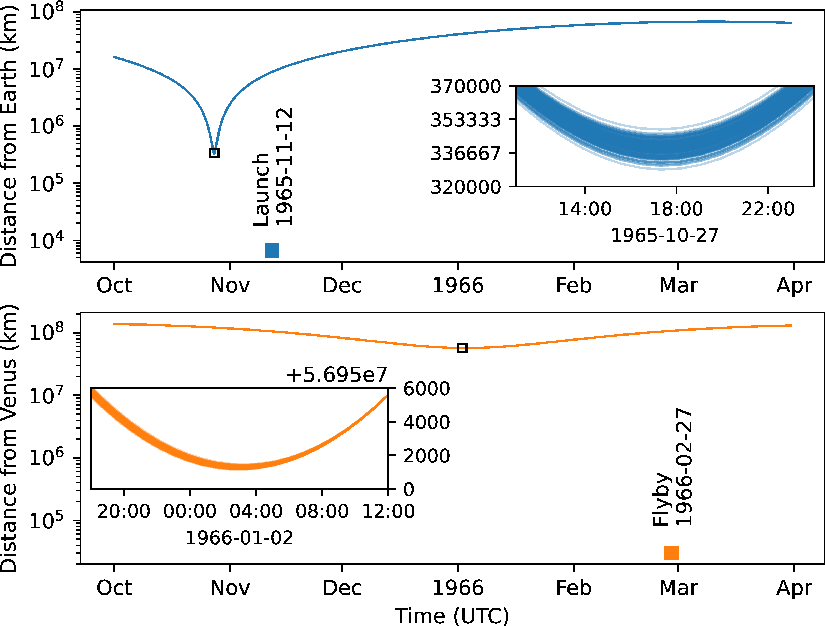}
\caption{Comparison between the orbit of \vl{} (solid lines) and the trajectory of Venera 2 at launch and Venus flyby (filled squares), in terms of the distance from the Earth and from Venus (top and bottom panel, respectively). For \vl{}, $180$ orbits are plotted, sampled from the normal distribution defined by the nominal orbital parameters and their covariance matrix obtained with \texttt{orbit\_finder} (since the lines overlap at the scale of the main panels, zoomed-in views around the closest approach to the Earth and to Venus are provided in the insets, to give a sense of the uncertainty). Reasonable values of the uncertainty in time and distance for the Venera 2 constraints are within the size of the square symbols.}
\label{default}
\end{center}
\end{figure*}

\section{Discussion and Conclusions}

To estimate how the uncertainty of my orbital solution of \vl{} affects the comparison in Figure 1, I used \texttt{emcee} \citep{emcee} to sample a multivariate normal distribution with mean equal to the nominal orbital parameters and covariance given by their covariance matrix obtained from \of{}.
Figure 1 shows $180$ of such sampled orbits, propagated through the time interval from 1965 Oct 10 to 1966 Mar 31. 
At the scale of the main panels, the sampled orbits are essentially indistinguishable, and the uncertainty on the orbit of \vl{} is within the thickness of the visible curve resulting from their overlap.
Similarly, for Venera 2, the conceivable uncertainty on the launch and flyby parameters (date and distance) is within the size of the square symbols.

Clearly, the orbit of \vl{} does not match either the launch or the flyby constraints by several orders of magnitude in distance, and by weeks, or more, in terms of date.
In particular, the minimum distance of \vl{} from the Earth in late 1965 is much larger than that of Venera 2 at injection, and also predates the launch by about fifteen days. 
The passage of \vl{} in the vicinity of Venus, which occurred in early 1966, is much shallower than the reported flyby range of Venera 2, and the mismatch in time is even larger (almost two months).

I conclude that the evidence available at this time does not support the identification of \vl{} with Venera 2 recently proposed by \citet{Loeb_Cloete:2025}.

\end{document}